# Plasmonic-dielectric compound grating with high group-index and transmission

**Lei Dai, Yang Liu, Chun Jiang[1,*]**

[1]*State Key Laboratory of Advanced Optical Communication Systems and Networks, Shanghai Jiao Tong University, Shanghai 200240, China*
[*]*cjiang@sjtu.edu.cn*

**Abstract:** We propose a compound system consisting of a dielectric grating and a plasmonic resonance cavity embedded in the grating. Based on the interference effect between the surface mode supported by the dielectric grating and the plasmonic-induced cavity mode, this system could achieve slow light with group index more than 200 and transmission more than 75%. Meanwhile, we examine the effects of the period numbers of the compound system and photonic crystal superlattice made up of alternate layers of the grating and air on the properties of slow light.

©2010 Optical Society of America

## 1. Introduction

Discovery of the extraordinary optical transmission (EOT) [1] through subwavelength nano-apertures perforated in metals has attracted a mass of experimental and theoretical studies devoting to interpreting the essential physics of the process in hole and slit arrays [2, 3, and the references therein]. There have been different mechanisms proposed to explain the physical origin of this EOT phenomenon depending on the geometrical parameters and wavelength. Let's take the studying of one-dimensional metal grating as an example. There are several possible explanations including the plasmonic cavity mode located inside the slits, the surface plasmonic modes on both surfaces of the grating [4], and the spoof surface waves of perforated perfect conductors [5]. Especially, the interference between these pathways may lead to the rich and complex transmission behavior, motivating the recent invention of many plasmonic-based photonic devices, such as nanoscale waveguides [6-8], hybridized plasmonic modes in nanoshells [9], plasmonic lasers [10-12] and switches [13], plasmonic-improved photovoltaic devices [14-15], and plasmonic focusing [16-17]. Meanwhile, the propagation of light pulses through subwavelength holes has also been studied to investigate the group delay of these subwavelength structures [18-19], where the silver film perforated with an array of subwavelength holes can achieve the effective group velocity of c/7 [18].

More importantly, in recent years the subwavelength plasmonic structure were exploited to mimic the electromagnetically induce transparency (EIT) [20-28] which is the quantum interference effect happening in atomic medium [29]. The basic idea of this plasmonic-induced EIT-like effect can be understood from two aspects. The first one is by introducing an EIT-like plasmonic "molecule" consisting of two plasmonic "atoms" supporting the radiative and dark states respectively [21]. The coupling between the radiative and dark states could induce the transparent window with strong dispersion and low absorption. The second one is by exciting "trapped mode" resonances [22, 30]. Actually, this trapped mode is not very unfamiliar, because it possesses the analogous mechanism and the asymmetric line shape with the Fano effect [31] occurring from interference between a direct and a resonance-assisted indirect pathways. Meanwhile, in contrast to the first mechanism in which the transparent window is based on the splitting of the absorption peak of the radiative state, the EIT-like effect based on trapped mode holds different feature where the absorption peak happens at the same wavelength with the transparent window. It is well known that the loss originates primarily from the metal and becomes more prominent when the concept of metamaterials-based devices is transferred to the optical part of the spectrum. Actually, this EIT-like effect has been realized theoretically and experimentally in many classical optical systems including the coupled silicon-ring resonator system [32-33] and the multiple coupled photonic-crystal cavity system [34-35], respectively.

In this paper, we explore a compound system to achieve low-loss slow light effect by integrating the transmission property of EOT and the mechanism of EIT, where the compound system consists of dielectric Si grating and periodically arranged plasmonic-induced resonance cavities. The waveguide mode supported by the Si grating is exploited as the direct pathway to interact with the resonance-assisted plasmonic cavity mode instead of the previous configurations where the direct pathway was also the plasmonic mode [22, 24]. Compared with the coupled system containing with two different plasmonic-induced resonance cavities, the proposed compound grating system can improve the efficiency of excited transparent window from less than 40% to more than 75% with group index more than 200. Next, the research of finite arrays effect of the compound system shows that the asymmetric transparent window can be shaped adequately by using the compound grating with only nineteen arrays of plasmonic cavities. Finally, we discuss the transmission spectra in 1D photonic-crystal superlattice made up of alternate layers of the designed compound grating and air. Apart from the regular radiative transparency and bandgap caused by Bragg resonance, there is a novel transparent window with high group index in the bandgap. Thus, we can use this multi-layer structure to increase the delay times. Meanwhile, we can conclude that the compound grating system may have potential applications of plasmonic sensing [36] and ultralow-power all-optical switching [37-38].

## 2. Results and Analysis

### 2.1. *Compound plasmonic grating design*

Fig.1 (a) shows the proposed plasmonic-dielectric compound grating structure consisting of alternate layers of silicon (n=3.5) waveguide and metal-dielectric-metal (MDM) plasmonic resonance cavity, where the plasmonic cavity is constructed by sandwiching $SiO_2$ ($n_1$=1.5) between two metal wires (Ag). We calculate the transmission and phase spectra of the compound grating using the finite-difference time-domain (FDTD) method [39]. The computational domain consists of a single period of the grating with resolution of 0.5nm. We use the periodic boundary condition along the periodic direction and the perfect matched layer (PML) absorption boundary condition at the top and bottom boundaries. We describe the complex optical constants of metal Ag taking from experimental data [40]. Throughout the article we only give attention to the plane wave with polarization along the x direction.

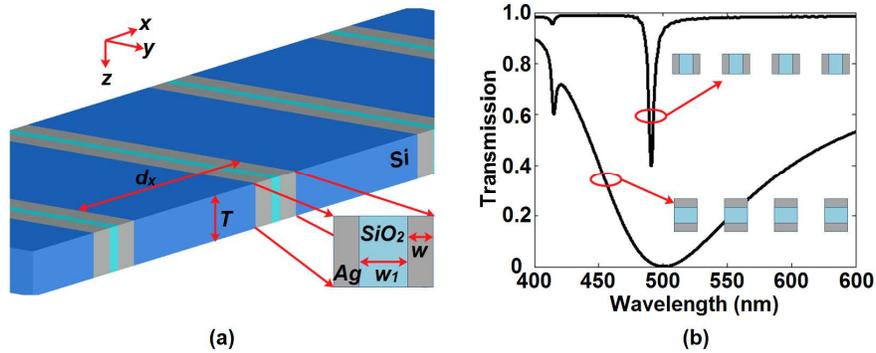

Fig.1 (a): Schematic of the compound plasmonic grating structure. The geometrical parameters are defined as in the figure: $d_x$ is the period and $T$ is the thickness of the Si waveguide. Light propagates along the $z$ direction with the polarization along the $x$ direction for TM mode. The enlarged picture is the metal-dielectric-metal plasmonic resonance cavity where the width of metal (w) and $SiO_2$ ($w_1$) are 10nm and 30nm respectively. (b): Simulated normal incidence transmission spectra of 1D periodic structures containing MDM plasmonic cavity or π/2-rotated plasmonic cavity

Before learning the properties of the compound grating structure, we begin by studying the periodic array of MDM plasmonic cavities as shown in Fig.1 (b). The period of the structure is 200nm; the width of metal and $SiO_2$ are 10nm and 30nm respectively; and the thickness of structure is 100nm. The result indicates that the MDM plasmonic cavity-constituted periodic structure holds two excited cavity modes, and these modes arise through the resonant excitation of the fundamental surface plasmonic polariton (spp) mode supported by the MDM waveguide. Meanwhile, truncation of such a MDM waveguide results in strong reflections from the terminations and gives rise to resonant cavity behavior. Actually, this MDM plasmonic cavity is a variation of the cut-wire pairs which is the basic element to construct the optical negative-index metamaterials [41-42]. The simulated transmission spectrum of this periodically arranged cut-wire pairs is also demonstrated in Fig.1 (b). From the results we can observe that this structure also supports two excited modes, and these modes are derived from the coupling between the metal wires holding the localized plasmonic polariton mode. This coupling results in the formation of two separated plasmonic eigenstates with opposite symmetric behavior. Usually the long-wavelength mode is the antisymmetric nature and is essential to provide negative permeability. Because the mechanism of the excited modes supported by the periodically arranged MDM cavities and cut-wire pairs is different, the ability to radiate the optical field also shows different characteristics. For example, the long-wavelength mode supported by the periodically arranged MDM cavities has bigger quality factor than the case of the periodically arranged cut-wire pairs. Thus, the excited modes supported by these periodic structures can be respectively considered as the resonance-assisted indirect pathway and the direct pathway based on the mechanism of Fano effect. Fig. 2(a) shows the coupled system incorporating the MDM plasmonic cavity and the cut-wire pairs in single period, and Fig.2 (b) demonstrates the simulated transmittance spectrum (red line) of this coupled system. The period of coupled system is 200nm, and the air gap between the MDM plasmonic cavity and the cut-wire pair is 9nm. The result indicates that there is a transparent window highlighted by shaded area at the wavelength 497nm. Next, the phase change and group index ($n_g$) of the coupled system are also illustrated in Fig.2 (c) in which the group index is calculated from the dispersion of the phase as follows:

$$n_g = \frac{c}{v_g} = \frac{c}{L}\tau_g = -\frac{c}{L}\frac{d\varphi(\omega)}{d\omega}$$

Here, $L$ is the thickness of the MDM cavity, $c$ is the speed of light in vacuum, $\tau_g$ is the delay time, and the phase $\varphi(\omega)$ is the function of the frequency $\omega$. We can observe from the phase spectrum that there are two frequency segments corresponding to the anomalous phase dispersion and one frequency segment corresponding to the normal phase dispersion around in the transparency window. Thus, we can draw the conclusion on the analogy of the EIT that the normal phase dispersion can result in slow light with the positive group index. Finally, the coupled system can achieve the maximum group index of 100 at the transmission peak. Meanwhile, the anomalous phase dispersion will demonstrate the negative group index corresponding to the opaque windows. In view of this negative group index we can understand from another formula of group velocity [43]:

$$v_g = \mathrm{Re}\left[\frac{d\omega}{dk}\right] = \frac{c}{\mathrm{Re}\left[n+\omega\,dn/d\omega\right]} \approx \frac{c}{n+\omega\,dn/d\omega}$$

$$n_g = n+\omega\,dn/d\omega$$

Here, n is the index of refraction which is the function of the frequency. Obviously, the negative group index is shaped when the index of refraction decreases rapidly enough with frequency. Meanwhile, it is well known that the refractive index takes a steep drop [44] around in the anomalous phase dispersion.

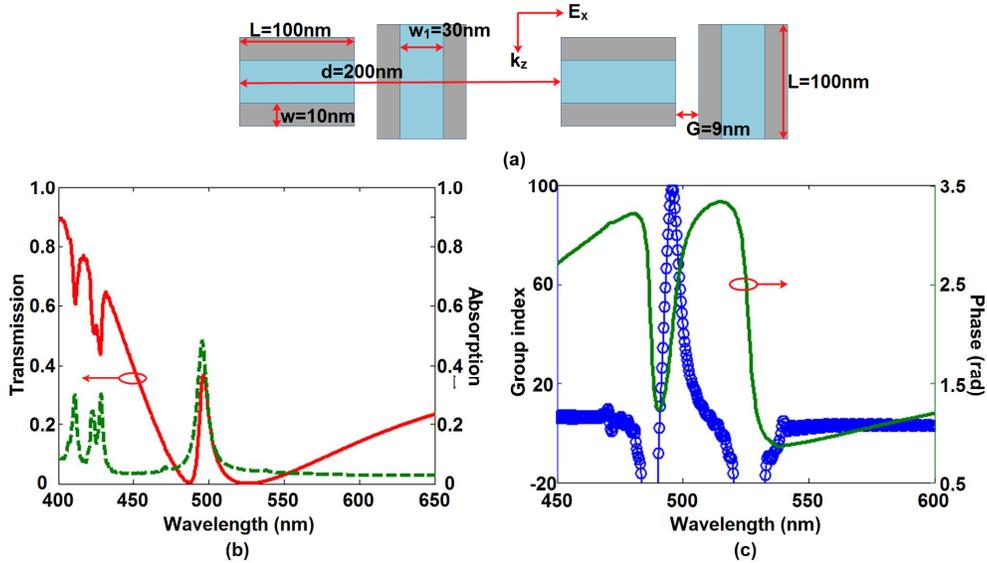

Fig.2: (a) Schematic of the coupled systems consisting of two MDM plasmonic resonance cavities separated by air gap G=9nm in single period d=200nm. The other geometrical parameters are defined as: the thickness of metal Ag w=10nm, the thickness of SiO2 $w_1$=30nm, the length of metal L=100nm. (b): Simulated normal incidence transmission spectrum (red) and absorption spectrum (green) of the coupled systems. (c): transmission phase change and calculated group index for the coupled systems.

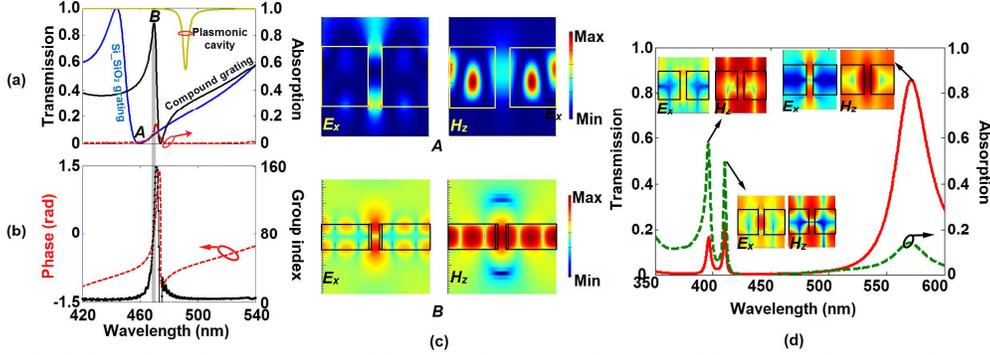

Fig. 3: (a) Normal incidence transmission (black) and absorption (red) spectra of the compound grating, where we also show the transmission spectra of Si grating filled with SiO$_2$ in the slits (blue) and plasmonic resonance cavity; (b) the corresponding phase change and group index ($n_g$) of the compound grating; the parameters for the compound grating in Fig.1 are used in the simulation as follows: period $d_x$=380nm; the thickness of grating T=100nm, the width of Ag w=10nm; and the width of SiO$_2$ $w_1$=30nm; (c): Electric $E_x$ and magnetic $H_z$ field intensity distributions at the wavelength locations A and B in (a); (d): Normal incidence transmission (red) and absorption spectra of metal grating filled with SiO$_2$ in the slits; the geometry parameters of metal grating are as follows: period d=380nm, grating thickness T=100nm, slits width w=50nm. Inset: electric and magnetic field intensity distributions of different resonance locations

Unfortunately, the transmission efficiency of the coupling system in Fig. 2 is very low, and the further analysis of the absorption spectrum in Fig.2 (b) reveals that the transmission peak and high absorption peak (>40%) occur at the same wavelength. The key reason for the high absorption is caused by the dissipation of metal when the optical energy is transported between the MDM plasmonic cavities and the cut-wire pairs. Thus, the compound grating system in Fig.1 (a) is proposed to reducing this loss, in which the low-loss Si grating is used as the direct pathway instead of the periodically arranged cut-wire pairs. Firstly, we begin by learning the resonance feature of the Si grating filled with SiO$_2$ in the slits of 50nm as shown in Fig. 3 (a), where the period of grating is 380nm and the thickness of grating is 100nm. The results indicate that this Si grating holds an excited mode with Fano asymmetric lineshape. After the resonance for this grating there is a dip with a transmission which completely vanishes. Through the observation of the electric and magnetic field distributions at this dip in Fig. 3 (c) we can understand that the formation of this dip is caused by the interference between the excited waveguide mode and the excited cavity mode in the slits. Thus, based on the mechanism of Fano effect, slow light effect can be achieved in this Si grating at the resonance. However, the group index for this Si grating is not large enough because the transmission phase change is not sharp. In order to achieve large group index the compound grating is proposed, and the normal-incidence transmission and absorption spectra of this compound grating is shown in Fig. 3 (a). We can observe clearly that an excited asymmetric transmission resonance with very steep lineshape is excited in this compound grating. This resonance is accompanied by a phase change of around π/2 (shaded area) and the corresponding group index calculated by this strong phase dispersion reaches 140 at the highest transmission peak as demonstrated in Fig. 3(b). From the electric field $E_x$ distribution at this peak in Fig.3 (c) we can observe that along the periodic direction there is an excited propagating surface mode that is concentrated on the silicon-air interface, and in the MDM cavity there is an excited spp-induced cavity mode. The asymmetric transparent effect is caused by the coupling between these modes. Although the waveguide surface mode decays exponentially away from the interface along the incidence direction, it can be radiated by coupling the energy into the plasmonic cavity to achieve the high transmission efficiency.

Meanwhile, because the dissipation of metal only happens in the MDM plasmonic cavity, this compound grating shows very small absorption as shown in Fig. 3 (a).

Next, we compare the compound grating of Fig.1 (a) with the metal grating filled with $SiO_2$ in the slits, in which the period and thickness of metal grating are 380nm and 100nm respectively, and the width of slits is 50nm. Fig.3 (d) shows the normal incidence transmission (red) and absorption spectra of this metal grating. From the results we can conclude that there are three resonant modes excited in this metal grating. From the electric and magnetic field distributions of these excited modes in the Figure we can observe that the long-wavelength excited mode with high transmission efficiency arises through the resonant excitation of the fundamental cavity mode located inside the slits, and the other two separated plasmonic modes are derived from the coupling between the localized plasmonic mode in the slits and the surface plasmonic modes on both surfaces of the grating. This coupling also can be explored to achieve the slow light effect, but the transmission efficiency is obvious lower than the case of the compound grating.

*2.2. Coupling strength discussion*

We can conclude that the thickness of the Ag metal layer plays a role in regulating the coupling strength between the dielectric surface mode and the plasmonic cavity mode. Thus, Fig.4 gives the influence of the thickness of the metal layer on the performances of the compound grating, where the structure parameters are same as Fig.3 (a). The results indicate that the transmission efficiency reduces with the increasing of the metal layer thickness, meaning that the coupling strength between the dielectric surface mode and the plasmonic cavity mode changes from strong to weak, and the asymmetric transparent window even disappears when the thickness of the metal layer is larger than the skin depth (~20nm). This property also can be understood from the field distributions as shown in Fig.4, where we depict the electric and magnetic field distributions corresponding to the high (A) and low (B) transparent windows respectively. Because the coupling strength becomes weak when the metal thickness increases, the dielectric surface mode radiates more difficultly through the plasmonic cavity than that of the strong coupling. Thus, the electric energy of the surface mode is confined mainly in the silicon waveguide. However, the slow-light effect shows the opposite property because the group index can be enhanced in the case of weak coupling as shown in Fig.4. Therefore, we can conclude that there is a mutual restraint between the transmission efficiency and group index. The comprehensive evaluation indicates that the compound system can achieve slow light with group index more than 200 and transmission efficiency more than 75%.

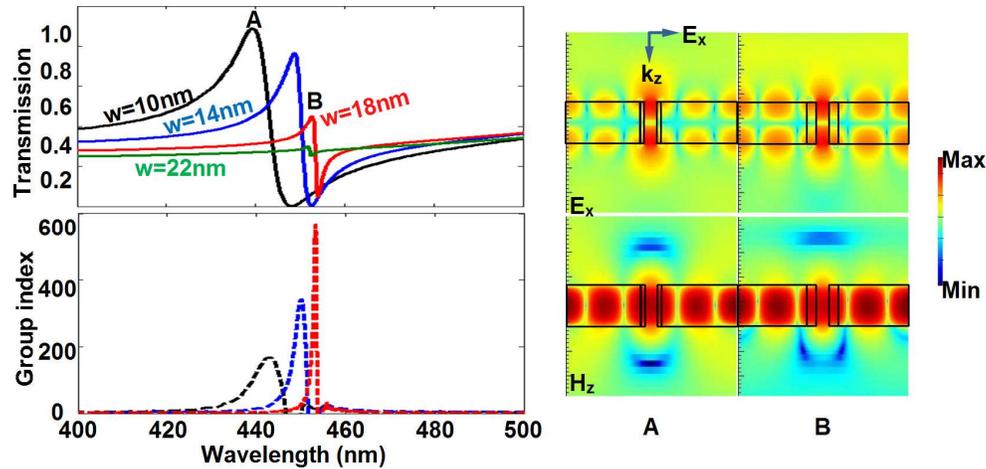

Fig.4: Normal-incidence transmission spectrums and the corresponding group index ($n_g$) calculated by the phase dispersion for the compound grating with the thickness of Ag metal layer increasing from 10nm to 22nm, where other structure parameters are used as follows: period $d_x$=380nm; grating thickness T=100nm, and $SiO_2$ width

$w_1$=30nm. The electric field $E_x$ and magnetic field $H_y$ are depicted corresponding to two different transparent windows.

*2.3. Finite-size effect*

Foregoing research is based on the compound grating system with infinite arrays of MDM plasmonic cavities in the x direction. However, in the actual study we should consider the dependence of transparency efficiency on the number of arrays. In other word, how many arrays in the grating are used to excite the surface waveguide mode coupled with the plasmonic-induced cavity mode? Fig.5 illustrates the calculated transmission spectrums for the compound grating with the number of arrays increasing from N=0 to N=19. The PML boundary condition is exploited in the numerical simulation instead of the periodic boundary condition along the periodic direction. The results reveal that the asymmetry transparent window is not excited when only one MDM plasmonic cavity is incorporated into the silicon waveguide, and the excited asymmetry transparent window becomes increasingly evident when the grating incorporates more MDM plasmonic cavities into the silicon waveguide. We can conclude that the asymmetric transparent window can be shaped adequately by using the system with only nineteen periods.

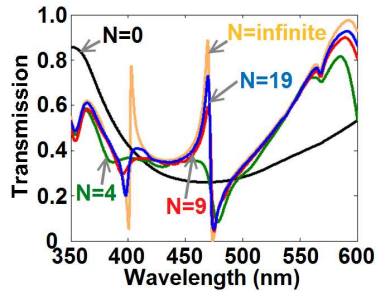

Fig.5: Normal incidence transmission spectrums for the compound grating with a finite number of arrays, where the period of grating $d_x$=380nm; grating thickness T=100nm, the width of Ag metal w=10nm; and SiO2 width $w_1$=30nm

*2.4. Photonic crystal superlattice*

In this section, we investigate one-dimensional photonic-crystal superlattice made up of alternate layers of the compound grating and air as shown in Fig.6 (a). The calculated transmission spectra of the photonic crystal superlattice with increasing number of layers are illustrated in Fig.6 (b), indicating that there are two band gaps and one bandwidth-enhanced transparent window caused by conventional Bragg resonance when the number of the layers increases to five, where in the first band gap there is a novel transparent window at wavelength of 470nm. Comparing with the transmission of the reference five-layer superlattice which is made up of alternate layers of the Si-air grating and air as shown in Fig.6 (c) (red line), we can understand that this novel transparent window is not caused by the Bragg resonance, but excited by coupling the radiated modes in plasmonic cavities between the fore-and-aft grating structures. Meanwhile, we can observe that although the transmission efficiency of the photonic crystal superlattice reduces from the 80% for the single-layer grating to only 35% for the five-layer structure, the group index of this five-layer structure shown in Fig.6 (d) remains at the high value of 100, meaning that the delay time can be increased by using this multi-layer superlattice.

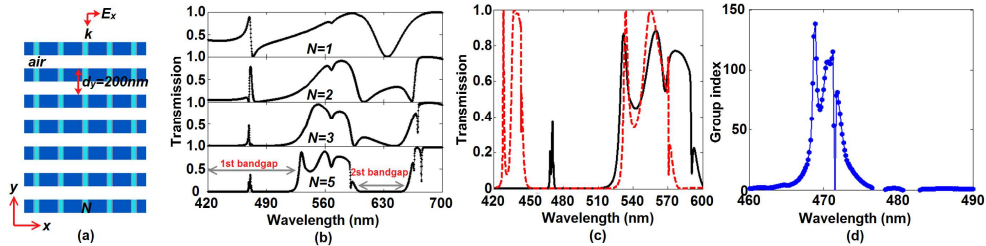

Fig.6: (a) the schematic of one-dimensional photonic crystal superlattice made up of alternate layers of the designed compound grating and air where the designed compound grating holds the same parameters in Fig.3 (a) and the thickness of air layer is 100nm. (b): the simulated transmission spectrums for the photonic crystal superlattice with the increasing of the number of the compound gratings. (c): the transmission spectrums of five-layer photonic crystal superlattice compared with the reference photonic crystal superlattice which is made up of alternate layers of the Si_air grating and air. (d): the corresponding group index of five-layer photonic crystal superlattice

## 3. Conclusion

In conclusion, we have explored a compound system of silicon grating structures which incorporate plasmonic-induced resonance cavities to achieve low-loss EIT-like effect. The results indicated that the compound grating could support the slow light with group-index more than 200 and transmission efficiency more than 75% in excited transparent window. Then, the investigation of the finite-period effect of the system revealed that the asymmetric transparent window could be shaped adequately by the compound grating with only nineteen arrays. Finally, we discussed the transmission properties in 1D photonic-crystal made up of alternate layers of the grating and air. Hence, the compound grating system may have potential for plasmonic sensing and ultralow-power all-optical switching.

This work was supported by National Natural Science Foundation of China (Grant No. 60672017) and sponsored by Shanghai Pujiang Program.